\documentclass[twocolumn,prb,groupedaddress,amsmath,amssymb,floatfix]{revtex4-1}
\usepackage{graphicx}
\usepackage{amssymb}
\usepackage{dcolumn}
\usepackage{bm}
\begin{document}

\title{Patchy worm-like micelles: solution structure studied by small-angle neutron scattering}

\author{Sabine Rosenfeldt,\textit{$^{a}$} 
Frank L\"{u}del,\textit{$^{b}$}  
Christoph Schulreich,\textit{$^{b}$} 
Thomas Hellweg,\textit{$^{b}$} 
Aurel Radulescu,\textit{$^{c}$} 
Joachim Schmelz,\textit{$^{d}$} 
Holger Schmalz\textit{$^{d}$} 
and Ludger Harnau\textit{$^{e,f}$}}

\email{harnau@fluids.mpi-stuttgart.mpg.de}

\affiliation{\textit{$^{a}$Physikalische Chemie I, Universit\"at Bayreuth, 
D-95440 Bayreuth, Germany}
\\\textit{$^{b}$Physikalische und Biophysikalische Chemie (PC III), Universit\"{a}t Bielefeld, 
D-33615 Bielefeld, Germany}
\\\textit{$^{c}$J\"ulich Centre for Neutron Science JCNS,
Forschungszentrum J\"ulich GmbH,
Outstation at FRM II,
Lichtenbergstra\ss e 1,
85747 Garching, Germany}
\\\textit{$^{d}$Makromolekulare Chemie II, Universit\"{a}t Bayreuth, 
D-95440 Bayreuth, Germany}
\\\textit{$^{e}$Max-Planck-Institut f\"{u}r Intelligente Systeme, 
Heisenbergstrasse 3, D-70569 Stuttgart, Germany}
\\\textit{$^{f}$Institut f\"{u}r Theoretische und Angewandte Physik, 
Universit\"{a}t Stuttgart, Pfaffenwaldring 57, 
D-70569 Stuttgart, Germany}
}
\date{\today}

\begin{abstract}
Triblock terpolymers exhibit a rich self-organization behavior including 
the formation of fascinating cylindrical core-shell structures with a phase separated corona. After 
crystallization-induced self-assembly of 
polystryrene-\textit{block}-polyethylene-\textit{block}-poly(methyl methacrylate) triblock 
terpolymers (abbreviated as SEMs = Styrene-Ethylene-Methacrylates) from solution,  worm-like 
core-shell micelles with a patchy corona of 
polystryrene and poly(methyl methacrylate) were observed by transmission electron microscopy. 
However, the solution structure is still a matter of debate. Here, we present a method to 
distinguish in-situ between a Janus-type (two faced) and a patchy (multiple compartments) 
configuration of the corona. To discriminate between both models the scattering intensity 
must be determined mainly by one corona compartment. Contrast variation in small-angle 
neutron scattering enables us to focus on one compartment of the SEMs. The results validate 
the existence of the patchy structure also in solution.
\end{abstract}
\maketitle

\section{Introduction}
Block copolymers exhibit a rich and fascinating self-assembly behavior in bulk 
and in selective solvents. \cite{foe97,bat99,ham01,hel11}  A lot of the occuring structures 
are promising for applications in drug delivery, optoelectronics or as scaffolds for 
nanoparticle assembly. \cite{ruz05,rod05,hab09,mot10,kim10a,kim10} Choosing the block 
length and the proper solvent conditions a huge variety of different structures can be 
generated. \cite{had05} 

Many of the solution-based assemblies can be summarized under the term multicompartment 
micelles. Similar to proteins, multicompartment micelles combine different physical 
nano-environments in well-segregated compartments and exhibit a rich phase behavior 
including remarkably complex self-assemblies. They show a compartmentalization either 
of the core or the corona. \cite{dup10,du11,zha12,mou12} Surface-compartmentalized particles 
exhibit useful features for several applications, e.g., the formation of hierarchically 
ordered superstructures, the use as potential scaffolds for the directed incorporation 
of metallic nanoparticles or as surfactants and emulsifiers. \cite{wal08a,wal08,voe:09,jia10,kre11}
Regarding corona-compartmentalized structures, Janus particles \cite{erh01,wal08a}
have been formed by template-assisted approaches while solution self-assembly mostly results 
in patchy particles, \cite{glo07,paw10,kre11} i.e., structures with more than two 
surface compartments. Whereas there are well-known examples for one-dimensional structures 
with compartmentalized cores \cite{dup10} or a Janus-like corona, \cite{liu03,ruh11} 
the majority of patchy particles are spherical in nature. Even though theoretical simulations 
by Binder et al. suggest the existence of one-dimensional nano\-struc\-tures with patch-like 
compartmentalization of the corona, \cite{the10,eru11,the11} only few examples have actually 
been published. For example, Liu et al. produced cylindrical micelles by dialysis of a triblock 
terpolymer against selective solvents that are able to further organize to double and 
triple helices. \cite{dup09}

In recent years, a new way of synthesizing stable anisotropic particles exploiting block 
copolymers with one crystallizable block moved into the focus of several research groups. 
\cite{laz09} Among these, polyferrocenylsilane containing block copolymers have been investigated 
most intensively, revealing a multitude of unprecedented structures, such as block co-micelles, 
scarf-like micelles and supramolecular brush layers. \cite{wan07,gae09,he11} The solution 
self-assembly of these crystalline-coil block copolymers is controlled by temperature or by the 
addition of a non-solvent for the crystallizable block, which induces crystallization. Especially, 
cylindrical or worm-like micelles with high aspect ratios have raised interest in bioscience and 
materials science. \cite{laz09,qia10}

Recently, we have developed the preparation of worm-like crystalline core micelles with a patchy 
corona from semi-crystalline polystyrene-\textit{block}-polyethylene-\textit{block}-poly(methyl
methacrylate) (SEM) triblock terpolymers in organic solvents. Our method provides a straightforward 
bottom-up strategy for building up one-dimensional patchy nanostructures via crystallization-induced 
self-assembly. The structure formation is triggered simply by a decrease in temperature that induces 
crystallization of the polyethylene (PE) middle block. \cite{schm:08c,schm:11} Transmission electron 
microscopy revealed that the corona exhibits a patchy structure made of microphase-separated 
polystyrene (PS) and poly(methyl methacrylate) (PMMA) enclosing the crystalline PE core. The complexity 
of surface-compartmentalized nanostructures complicates the determination of the morphology and 
dimensions. Up to now, morphological information is obtained by imaging techniques that usually 
were applied to dried samples. To get a deeper insight in dissolved worm-like crystalline core micelles, 
scattering methods such as small-angle neutron scattering are powerful tools as has been shown 
for worm-like or Janus-type structures by F\"utterer et al. \cite{fue04} and by Walther et al., 
\cite{walt:09} respectively. Here, we present the first in-situ shape sensitive investigation of patchy 
worm-like micelles from a SEM triblock terpolymer. To achieve this goal, a theoretical model for these 
complex structures is developed and experimentally verified by small-angle neutron scattering  
on  patchy worm-like crystalline core micelles containing deuterated polystyrene blocks (\textit{d}SEM) 
at a selected contrast.

\section{Experimental section}
\subsection{Synthesis and sample preparation}

The $d$SEM triblock terpolymer was obtained by catalytic hydrogenation of the 
corresponding $d$SBM (B$=$poly(1,4-butadiene)) triblock terpolymer precursor 
synthesized by sequential anionic polymerization. The polystyrene block of the 
precursor was fully deuterated. The composition of the $d$SBM precursor is 
determined to be S$_{\mathrm{47}d}$B$_{\mathrm{24}}$M$_{\mathrm{29}}^{\mathrm{85}}$ 
by a combination of MALDI-TOF and $^{\mathrm{1}}$H-NMR, which results in 
S$_{\mathrm{47}d}$E$_{\mathrm{24}}$M$_{\mathrm{29}}^{\mathrm{86}}$ after 
hydrogenation (subscripts denote the mass fraction in percent, the superscript 
gives the overall molecular weight in kg/mol, and \textit{d} indicates that the PS 
block is fully deuterated). 
The formula of the
investigated
S$_{\mathrm{47}d}$E$_{\mathrm{24}}$M$_{\mathrm{29}}^{\mathrm{86}}$ can also be expressed in terms of
the number of monomer units and would read S$_{\mathrm{359}d}$E$_{\mathrm{747}}$M$_{\mathrm{250}}$.
Full saturation of the double bonds was confirmed by  
$^{\mathrm{1}}$H-NMR in deuterated toluene at 65$^{\rm o}$C. A detailed description 
of the synthesis of the SEM terpolymer is given in the literature. \cite{schm:08c} 
Micelles of S$_{\mathrm{47d}}$E$_{24}$M$_{29}^{86}$ are formed by crystallization 
induced self-assembly upon cooling. \cite{schm:08c} As the polyethylene block in 
a 10 g/L solution melts at a peak melting temperature $T_{m} = 45^{\rm o}$C and 
crystallizes at $T_{c} = 21^{\rm o}$C, the solutions for the scattering experiments 
were prepared as follows: To eliminate any influence of thermal history, 10 g/L of 
the $d$SEM were dissolved in the corresponding solvent, e.g., in tetrahydrofuran 
(THF) or a mixture of protonated and deuterated THF (deuteration degree 99.5{\%}, 
Deutero GmbH) at 65$^{\rm o}$C. After 1 h the solutions were quenched down to 
20$^{\rm o}$C in a water-bath and equilibrated for two days.

\subsection{Small angle neutron scattering (SANS)}

The SANS data were obtained using the KWS 1 instrument at the FRM II in 
Munich, Germany. The raw data were 
corrected for background, solvent and empty cell scattering
by the use of the software provided by the J\"ulich Center for Neutron
Science (JCNS) at the FRM II. Absolute intensities were obtained by
using a calibrated reference scatterer. For all 
data sets, the rate of incoherent scattering caused by the protons was 
determined at high scattering vector, set as a constant and subtracted from 
the raw data. 

\subsection{Transmission electron microscopy (TEM)}

For TEM studies the solutions were diluted to 1 g/L. Samples were prepared by 
placing a drop of the solution on a carbon-coated copper grid. After 20 s, excess 
solution was removed by blotting with a filter paper. Subsequently, elastic 
bright-field TEM was performed on a Zeiss 922 OMEGA EFTEM (Zeiss NTS GmbH, 
Oberkochen, Germany) operated at 200 kV. Staining was performed with 
RuO$_{\mathrm{4}}$ vapor for at least 20 min. RuO$_{\mathrm{4}}$ is known to 
selectively stain PS, which enables to distinguish between PS and PMMA domains 
in the corona of the micelles.

\section{Results and Discussion}
\subsection{TEM micrographs}

\begin{figure}[t!]
\centering
\includegraphics[width=8cm]{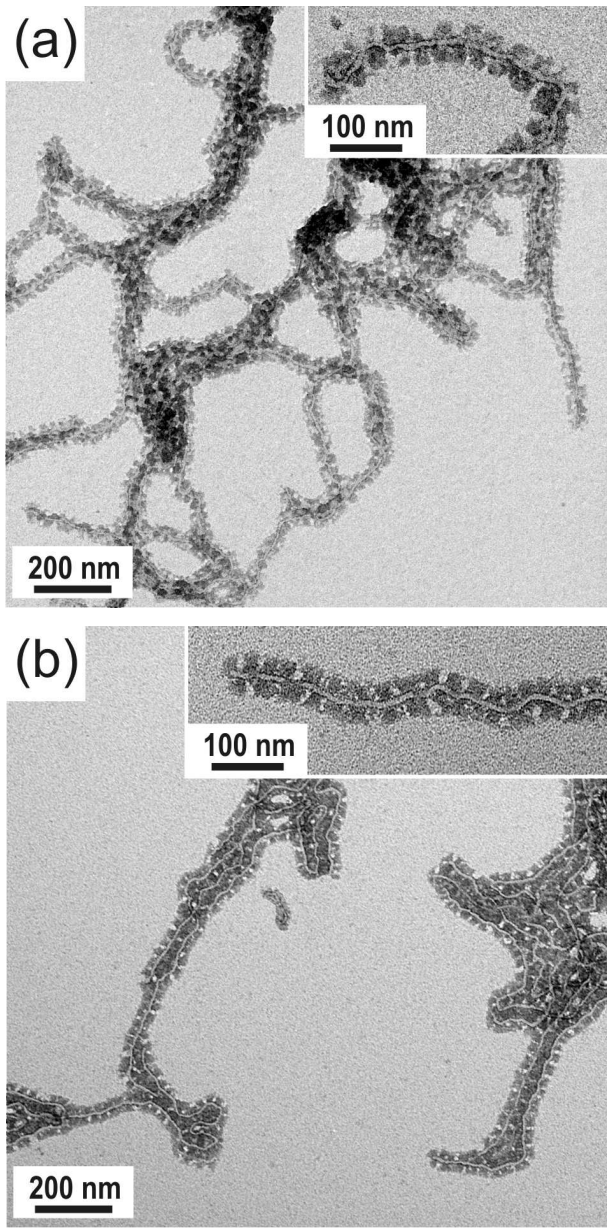}
\caption{TEM micrographs of the self-assembled structure of 
S$_{\mathrm{47}d}$E$_{24}$M$_{29}^{86}$ (1 g/L) in THF (a) and deuterated THF (b).
The PS domains are stained with RuO$_{4}$ and therefore visible as dark gray areas.
}
\label{fig1} 
\end{figure}

Recently, the formation of worm-like crystalline core micelles with a patch-like 
corona from a polystryrene-\textit{block}-polyethylene-\textit{block}-poly(methyl 
methacrylate) triblock terpolymer was reported. \cite{schm:08c,schm:11}
2D $^{\mathrm{1}}$H-NMR NOESY techniques applied to a 
S$_{\mathrm{39}}$E$_{\mathrm{21}}$M$_{\mathrm{40}}^{\mathrm{91.5\thinspace 
}}$ in toluene pointed to a micro phase-separation of the corona. \cite{schm:08c} 
However, this technique is not able to distinguish between a Janus-type 
(two-faced) or a patchy (multiple PS and PMMA compartments) configuration of 
the corona. Hence, the assumption of a patchy worm-like structure in solution was based
on TEM studies. Fig.~\ref{fig1} shows a TEM micrograph of the structures formed by 
quenching a solution of S$_{\mathrm{47}d}$E$_{\mathrm{24}}$M$_{\mathrm{29}}^{\mathrm{86}}$ 
from 65$^{\rm o}$C to 20$^{\rm o}$C. In order to distinguish the different compartments 
in the corona in the dried state, the PS domains were stained with RuO$_{\mathrm{4}}$. 
The TEM micrographs clearly exhibit similar patch-like compartments of the corona both 
in THF (see Fig.~\ref{fig1}(a)) and in deuterated THF (see Fig.~\ref{fig1}(b)). Hence, 
a change in the corona structure due to isotope effects of the solvent can be 
ruled out. In many sections an alternating array of the PS patches along the 
core of the worm-like crystalline core micelles is observed. 
THF is a good solvent for both PS and PMMA  and the adopted random-coil
configuration of the chains results in different dimensions of the 
hemi-shells. A detailed discussion 
about TEM and cryo-TEM studies on worm-like crystalline core micelles formed by SEM 
terpolymers can be found elsewhere. \cite{schm:08c,schm:11}
%
%
%
%

\subsection{Solution structures as obtained from SANS}

Scattering techniques provide knowledge about the solution structure without perturbing 
the sample. Moreover, in the case of neutron scattering, contrast variation using 
deuterated monomers reveals details of the internal structure in a unique way. 
SANS data taken at highest contrast between solvent and solute are used to explore 
the shape of the entire species. At intermediate and low contrast local details of 
the self-assembled structures can be detected. The scattering intensities of such a 
contrast series can be interpreted by applying models with appropriate geometry and 
scattering length density distribution. \cite{will:91,bals:94,higg:96,rose:09,walt:09}

\subsubsection{Scattering intensity}

SANS determines the scattering intensity $I({\bf q})$ as a function of the 
scattering vector ${\bf q}$ and the concentration of the dissolved particles. 
In addition to the coherent scattering intensity $I_{coh}({\bf q})$, there is always an 
incoherent contribution $I_{incoh}$ that is due to the protons present in the 
particles under consideration. The scattering intensity can be written as 
\begin{equation} \label{eq1}
I({\bf q}) = I_{coh}({\bf q}) + I_{incoh}\,.
\end{equation}
Note that in the notation the dependence on the concentration of the dissolved particles 
is suppressed. The ${\bf q}$-independent incoherent contribution $I_{incoh}$ of individual 
particles must be subtracted carefully from experimental data in order to obtain 
meaningful results on the structure and interaction of the dissolved particles. 
\cite{higg:96} Due to the mesoscopic scale of the particles, the solvent will be 
modeled as structureless continuum providing a homogeneous scattering length density
$b_{solvent}$.

In order to take into account particle polydispersity  we consider a multicomponent 
system involving $\nu$ species of particles with particle numbers $N_O$ ($1 \le O \le \nu$) 
in the volume $V$. Each particle of a species $O$ carries $n_O$ scattering units. In the 
case of the triblock terpolymer micelles under consideration, it proves convenient to assign 
an index $j$ ($1 \le j \le n_O$) to scattering units, and to order them such that units 
$1 \le j \le n_O^{(E)}$ belong to the compound PE, $n_O^{(E)} + 1 \le j \le n_O^{(E)} 
+ n_O^{(M)}$ belong to the compound PMMA, and $n_O^{(E)} + n_O^{(M)}+ 1 \le j \le n_O$
belong to the compound PS. The coherent contribution to the scattering intensity in the 
$\nu$-component system is given by
\begin{equation}  \label{eq2}
I_{coh}({\bf q}) = \sum\limits_{O,P=1}^\nu  I_{OP}({\bf q})\,,
\end{equation}
with the partial scattering intensities 
\begin{eqnarray}
\lefteqn{I_{OP}({\bf q}) = } \nonumber
\\&&\frac{1}{V}\left\langle
\sum\limits_{j=1}^{n_O} \sum\limits_{k=1}^{n_P}
\sum\limits_{\alpha=1}^{N_O} \sum\limits_{\gamma=1}^{N_P}
\Delta {\tilde b}_{jO}^{(\alpha)} \Delta {\tilde b}_{kP}^{(\gamma)}
e^{i{\bf q}\cdot \left({\bf r}_{jO}^{(\alpha)}-{\bf r}_{kP}^{(\gamma)}\right)} \label{eq3} 
\right\rangle\,.
\end{eqnarray}
Here,  ${\bf r}_{jO}^{(\alpha)}$  is the position vector of the $j$-th scattering 
unit of the $\alpha$-th particle of species $O$. The difference of the scattering 
length of this scattering unit and the average scattering length of the solvent is 
denoted as $\Delta {\tilde b}_{jO}^{(\alpha)}$, and $\langle \,\,\,\, \rangle$ 
denotes an ensemble average. It proves convenient to decompose the partial scattering 
intensities according to 
\begin{equation}   \label{eq4}
I_{OP}({\bf q}) = \rho_O \omega_O({\bf q})\delta_{OP}
+ \rho_O \rho_P h_{OP}({\bf q})\,,
\end{equation}
where
\begin{eqnarray}
\lefteqn{h_{OP}({\bf q}) = } \nonumber
\\&&
\frac{V}{N_ON_P}\left\langle
\sum\limits_{j=1}^{n_O} \sum\limits_{k=1}^{n_P}
\sum\limits_{\alpha=1}^{N_O} \sum\limits_{\stackrel{\gamma=1}{\gamma\neq \alpha}}^{N_P}
\Delta {\tilde b}_{jO}^{(\alpha)} \Delta {\tilde b}_{kP}^{(\gamma)}
e^{i{\bf q}\cdot \left({\bf r}_{jO}^{(\alpha)}-{\bf r}_{kP}^{(\gamma)}\right)}
\right\rangle\,\,\label{eq5} 
\end{eqnarray}
is a particle-averaged total correlation function for pairs of particles of species 
$O$ and $P$. The number density of particles of species $O$ is designated as 
$\rho_O = N_O /V$. The particle-averaged intramolecular correlation function
\begin{equation}  \label{eq6} 
\omega_O({\bf q}) = \frac{1}{N_O}\left\langle
\sum\limits_{j=1}^{n_O} \sum\limits_{k=1}^{n_P}
\sum\limits_{\alpha=1}^{N_O}
\Delta {\tilde b}_{jO}^{(\alpha)} \Delta {\tilde b}_{kO}^{(\alpha)}
e^{i{\bf q}\cdot \left({\bf r}_{jO}^{(\alpha)}-{\bf r}_{kO}^{(\alpha)}\right)}
\right\rangle
\end{equation}
characterizes the scattering length distribution, and hence also the geometric shape 
of particles of species $O$. While the particle-averaged intramolecular correlation 
functions account for the interference of radiation scattered from different parts of 
individual particles in a scattering experiment, the local order in the fluid is characterized 
by the total correlation functions. For flexible particles the intramolecular correlation 
functions depend on the particle number density and follow from a statistical average 
over particle configurations. As suggested by the imaging data (see Fig.~\ref{fig1}), the 
tribock terpolymer micelles may be considered as worm-like core-shell cylinders with 
phase-separated shells. In the scattering vector regime of a SANS experiment both the 
contour length and the persistence length cannot be resolved for rather long and stiff 
cylindrical particles \cite{higg:96,yan:08}.
As a prerequisite for the following analysis we have confirmed that the scattering intensity
of a homogeneous weakly bendable cylinder with the same configuration as the 
triblock terpolymer micelle shown in the inset of Fig.~\ref{fig1}(b) is nearly 
indistinguishable from the scattering intensity of a corresponding homogeneous rigid cylinder 
in the scattering vector regime accessible by SANS.
Hence, only the scattering intensity of rigid 
cylinders is considered in the analysis of the SANS experiments. In the limit of a continuous 
distribution of scattering units, the intramolecular correlation function of randomly oriented 
core-shell cylinders is given by 
\begin{equation}  \label{eq7}
\omega_O(q) = \frac{1}{4 \pi}\int\limits_0^{2\pi}d\phi\,\int\limits_0^\pi d\theta\,
\sin\theta\,F_O(q,\theta,\phi) F_O^\star(q,\theta,\phi)
\end{equation}
with
\begin{equation}  \label{eq8}
F_O(q,\theta,\phi) = \int\limits_{-L/2}^{L/2} dz\, \int\limits_0^\infty dr\,r
\int\limits_0^{2\pi} d\phi_r\,\Delta b_O(r,\phi_r,z)e^{i {\bf q}\cdot {\bf r}}\,.
\end{equation}
Here, ${\bf q}=q (\sin\theta \cos\phi, \sin\theta \sin\phi, \cos\theta)$ is the 
scattering vector described by spherical coordinates and ${\bf r}=(r \cos\phi_r, 
r \sin\phi_r, z)$ denotes the position vector of a scattering unit of an individual 
core-shell cylinder described by cylindrical coordinates. The origin of the coordinates
is taken to be the center of the cylinder of length $L$ and $\Delta b_O(r,\phi_r,z)$ 
denotes the scattering length density function, which specifies the internal structure 
of an individual core-shell cylinder. For the triblock terpolymer micelles the index $O$  
in eqn (\ref{eq2}), (\ref{eq4}), (\ref{eq7}), and  (\ref{eq8}) allows one to take into 
account polydispersity of the scattering length density function.

\subsubsection{Janus type and patchy cylinders}

\begin{figure}[t!]
\centering
\includegraphics[width=8cm]{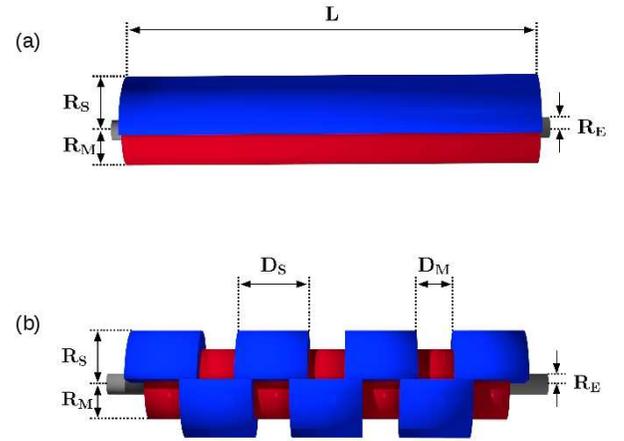}
\caption{Illustrations of two possible architectures of micro phase-separated shells 
of cylindrical particles. The cylinders are characterized by a core-shell structure 
with a PE core of radius $R_E$ marked in gray and a biphasic PS-PMMA shell consisting
of regions of unlike size and scattering length densities marked in blue and red. In 
the main text the cylindrical particles are denoted as Janus cylinders (a) and patchy 
cylinders (b).}
\label{fig2} 
\end{figure}

\begin{figure}[ht!]
\begin{center}
\includegraphics[width=7.5cm,clip]{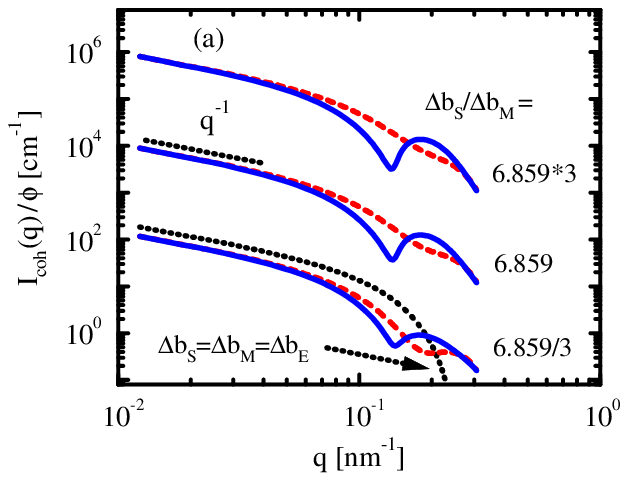}\\[15pt]
\includegraphics[width=7.5cm,clip]{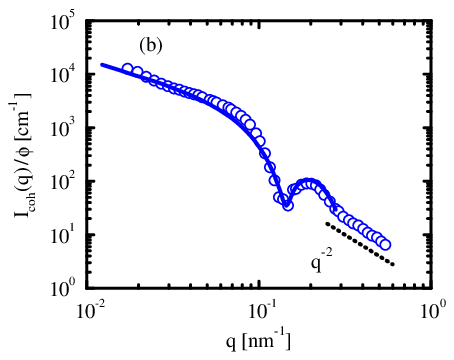}
\caption{
(a) Comparison of the scattering intensity $I_{coh}(q)/\phi$ of Janus cylinders 
(dashed lines, see Fig.~\ref{fig2}(a)) with the scattering intensity of patchy 
cylinders (solid lines, see Fig.~\ref{fig2}(b)) with $D_M=D_S=22.4$ nm. 
The curves have been calculated according to eqn (\ref{eq1}) - (\ref{eq8}) with 
$h_{OP}({\bf q})=0$ and $\nu = 1$, i.e.,  for noninteracting and monodisperse 
cylinders. The remaining model parameters are given by $L=425.5$ nm, 
$R_E=4.3$ nm, $R_M=15.5$ nm, $R_S=22.0$ nm, $\Delta b_S/\Delta b_E = -0.751$ and 
$\Delta b_E = 0.52\times 10^{10}$ cm$^{-2}$. Moreover, the ratio of the scattering 
length density difference of PS and PMMA decreases from top to bottom according to
$\Delta b_S/\Delta b_M = 6.859*3, 6.859, 6.859/3$, where the value 6.859 corresponds 
to the actual experimental system discussed in panel (b). In addition the lower 
dotted line depicts the scattering intensity of a homogeneous 
cylinder with $\Delta b_S/\Delta b_E =\Delta b_S/\Delta b_M = 1$ and $R_M=R_S=15.5$ nm.
For clarity, the upper and lower data sets have been shifted up and down, respectively, 
by a factor of 10$^2$. 
(b) Measured scattering intensity of the triblock terpolymer micelles (10 g/L) in 
THF$_{\rm H}$ (symbols) together with the calculated results for noninteracting 
patchy cylinders (solid line). The model parameters are the same as for the solid 
middle line in panel (a) except of $D_M=14.0$ nm and $D_S=29.0$ nm. The short dotted 
lines in panels (a) and (b) represent two asymptotic scaling laws as discussed in the 
main text.}
\label{fig3}
\end{center}
\end{figure}

As Fig.~\ref{fig2} illustrates, the cylinders are characterized by a core-shell 
structure with a PE core of radius $R_E$ marked in gray and a biphasic PS-PMMA shell 
consisting of regions of unlike size and scattering length densities shown in blue 
and red. Hence the scattering length density function equals $\Delta b_E$, $\Delta b_S$, 
and $\Delta b_M$ in the regions marked in gray, blue, and red, respectively. More 
specifically, two shell patterns will be distinguished. (i) Fig.~\ref{fig2}(a): 
A Janus-type architecture, i.e., the shell consists of two homogeneous hemi-shells which
might have different or similar extensions ($R_S$ and $R_M$) 
and unlike scattering length densities. (ii) Fig.~\ref{fig2}(b): The two 
inhomogeneous hemi-shells consist of alternating regions of scattering length densities. 
Here, $D_S$ and $D_M$ describe the lengths of the alternating regions of the so-called 
patchy cylinder. 
%
%
%
%
Fig.~\ref{fig3}(a) displays scattering intensities for noninteracting 
($h_{OP}({\bf q})=0$) and monodisperse ($\nu = 1$) cylinders of length \mbox{$L=425.5$ nm} 
and radii \mbox{$R_E=4.3$ nm}, \mbox{$R_M=15.5$ nm}, \mbox{$R_S=22.0$ nm} as 
calculated from eqn (\ref{eq1}) - (\ref{eq8}). The dashed lines show the results for 
Janus cylinders (see Fig.~\ref{fig2}(a)), while the solid lines represent the scattering 
intensities of patchy cylinders (see Fig.~\ref{fig2}(b)) with \mbox{$D_M = D_S = 22.4$ nm}. 
All scattering intensities are normalized to 
the volume fraction $\phi=\rho_O/V_O$, where $V_O$ is the particle volume. 
Moreover, the ratio of the scattering length density differences of 
PS and PE is $\Delta b_S/\Delta b_E = -0.751$, where $\Delta b_E = b_{solvent} - b_E$ 
and $b_E = -0.34\times 10^{10}$ cm$^{-2}$ are the scattering length densities of the 
solvent and PE, respectively. For comparison, three values of the ratio of the scattering 
length density differences of PS and PMMA $\Delta b_S/\Delta b_M$ are 
considered, where $\Delta b_S/\Delta b_M = 6.859$ (data set in the middle) corresponds 
to PS and PMMA in THF$_{\rm H}$. The scattering intensities 
shown in Fig.~\ref{fig3}(a) have the following features. They exhibit the $q^{-1}$ scaling 
relation (short dotted line) for small scattering vectors which is characteristic for the 
linear arrangement of scattering units along the main axis of a cylinder. For the patchy 
cylinders (solid lines), the scattering intensity exhibits a minimum at 
\mbox{$q \approx 0.14$ nm$^{-1}$}, while for the Janus cylinders no minimum is observed, 
provided $\Delta b_S/\Delta b_M \gtrsim 6.859/3$. From the figure it is apparent that 
the existence or absence of a minimum of the scattering intensity at intermediate scattering 
vectors allows one to distinguish Janus cylinders from patchy cylinders provided the ratio of 
two scattering length density differences of the biphasic shell is sufficiently large. 
In addition we note that the scattering intensity of a homogeneous cylinder of similar size 
does not exhibit a minimum at \mbox{$q \approx 0.14$ nm$^{-1}$} as is apparent from the 
lower dotted line in Fig.~\ref{fig3}(a).

%
%
%
%
In Fig.~\ref{fig3}(b) the experimental scattering intensity of the triblock 
terpolymer micelles in pure THF$_{\rm H}$ is compared to the calculated results for 
noninteracting patchy cylinders. For the morphological study this contrast
condition is well suited due to the fact that the scattering pattern mainly is determined
by the deuterated PS patch of the shell.
The model parameters are the same as for the solid middle line in
Fig.~\ref{fig3}(a) except of the lengths of the alternating regions of the 
shell which are given by \mbox{$D_M=14.0$ nm} and \mbox{$D_S=29.0$ nm}. Hence we take 
into account that the size of the PMMA block (M$_{\mathrm{29}}$) characterized by $D_M$, 
$R_M$ is smaller than the size of the PS block (S$_{\mathrm{47}d}$) characterized by $D_S$, 
$R_S$. The ratio of the size of the three-dimensional patches $D_S (R_S^2-R_E^2) /
(D_M (R_M^2-R_E^2))=4.35$ is similar to the cube of the ratio of the mass fractions 
of the one-dimensional PS and PMMA chains $(47/29)^3=4.26$. From the figure it is 
apparent that the experimentally determined scattering 
intensity (symbols) exhibits a pronounced minimum at \mbox{$q \approx 0.14$ nm$^{-1}$} 
which is characteristic for a patchy cylinder as discussed above. The deviations between 
the experimental data and the calculated results (solid line) for intermediate scattering 
vectors $q \in [0.03,0.1]$ are due to the fact that the PS and PMMA blocks do not form 
perfect patchy half-cylinders as is assumed in the model shown in Fig.~\ref{fig2}(b). 
Nevertheless, the combination of Figs.~\ref{fig3}(a) and (b) demonstrates that 
patchy triblock terpolymer micelles are indeed present in THF$_{\rm H}$. For comparison 
we emphasize that experimentally determined scattering intensities of Janus cylinders 
do not exhibit minima at intermediate scattering vectors in agreement with the calculated 
results shown in Fig.~\ref{fig3}(a). \cite{walt:09}

Although eqn (\ref{eq7}) and (\ref{eq8}) have been derived for a rigid cylinder of a 
definite shape illustrated in Fig.~\ref{fig2}, in reality concentration fluctuations of 
the PS and PMMA polymer chains contribute to the scattering intensity. On the basis 
of our experience with various polymer nanoparticles \cite{henz:08,yan:08,roch:11,henz:11}
we expect that the contribution of the polymer concentration fluctuations becomes important 
for large scattering vectors \mbox{$q \gtrsim 0.3$ nm$^{-1}$}. Within a Gaussian approximation 
the scaling relation $I_{coh}(q) \sim q^{-2}$ is valid for large scattering vectors as 
indicated by the dotted line in Fig.~\ref{fig3}(b). Moreover, we have confirmed that 
moderate size polydispersity (e.g., \mbox{$D_S=29.0 \pm 3.0$ nm}) does not lead to 
pronounced changes of the calculated scattering intensity. Therefore, one may consider 
a monodisperse model system as an appropriate approximation.

\subsubsection{Contrast variation}

%
%
\begin{figure}[t!]
\centering
\includegraphics[width=7.5cm,clip]{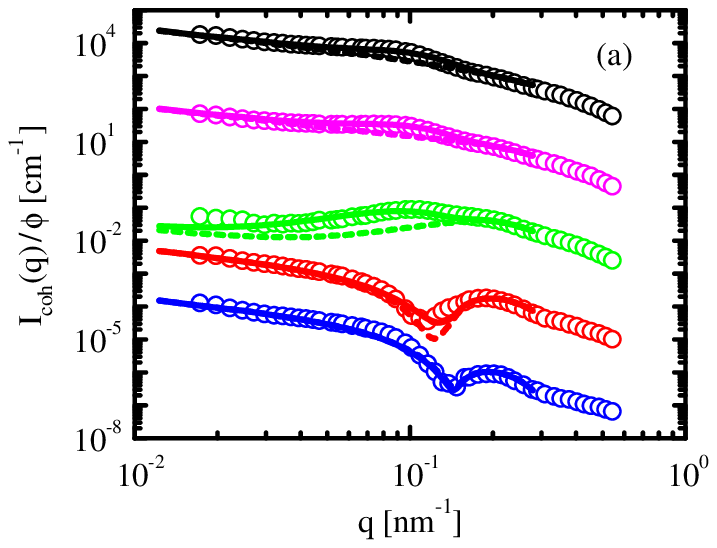}\\[15pt]
\includegraphics[width=7.5cm,clip]{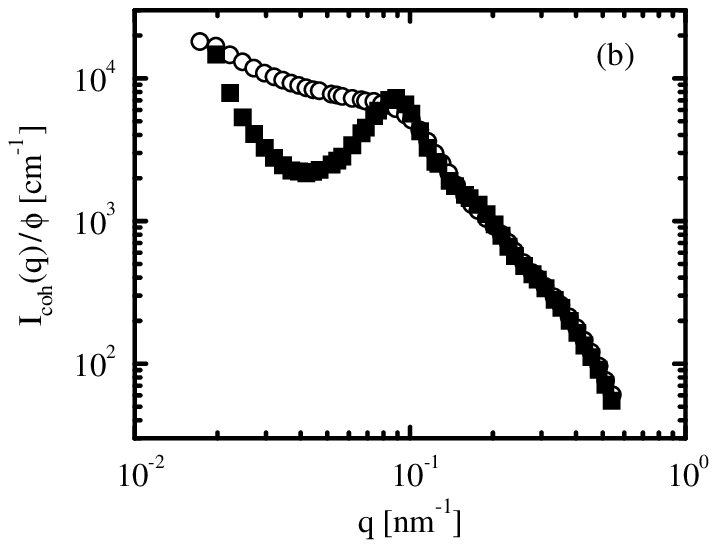}
\caption{(a) Measured scattering intensity $I_{coh}(q)/\phi$ of the triblock terpolymer 
micelles at concentration 10 g/L (symbols).
The scattering length densities of the solvent increases from bottom to top 
($b_{solvent} = 0.18, 1.36, 3.23, 5.13, 6.32 \times 10^{10}$ cm$^{-2}$) while the 
corresponding ratios of the scattering length density differences of the micelles are 
given by 
$\Delta b_S/\Delta b_M = 6.859, -19.731, -1.531, -0.337, -0.033$ and 
$\Delta b_S/\Delta b_E =-0.751,  -0.187, -0.057, -0.015, -0.002$, with
$\Delta b_E = b_{solvent} - b_E$ and $b_E = -0.34\times 10^{10}$ cm$^{-2}$. The 
four lower scattering intensities are shifted down by a factor of 10$^2$, 10$^4$, 
10$^6$, 10$^8$, respectively. The dashed lines represent calculated results for 
noninteracting ($h_{OP}({\bf q})=0$) patchy cylinders using the same model parameters 
as in Fig.~\ref{fig3}(b). In the case of the solid lines a contribution of the 
total correlation function $h_{OP}({\bf q})$ to the scattering intensity is taken 
into account. For the lowest scattering length density of the solvent the solid 
line is nearly indistinguishable from the dashed line.
(b) Comparison of the normalized measured scattering intensity of the
triblock terpolymer micelles at concentration 10 g/L (open circles) with the 
corresponding data at 40 g/L (solid squares) in fully deuterated THF 
($b_{solvent} =  6.32 \times 10^{10}$ cm$^{-2}$). The differences between the open 
and solid symbols reflect pair correlations between the micelles.}
\label{fig4} 
\end{figure}
%
%

Fig.~\ref{fig4}(a) displays SANS intensities of the triblock terpolymer micelles 
in different THF$_{\rm D}$:THF$_{\rm H}$ mixtures corresponding to different scattering 
length densities of the solvent (symbols). Such a contrast variation allows 
consistency checks of the theoretical modeling because the contribution of the three 
polymers PE, PMMA, and PS to the scattering intensity depends sensitively on the 
scattering length density of the solvent. The figure demonstrates that varying the 
scattering length contrast leads to marked differences in the scattering intensities. 
In particular, the minima of the scattering intensities of the lower two data sets 
disappear upon increasing the scattering length density of the solvent 
(upper three data sets) due to an increasing contribution of PE to the scattering 
intensity. The pronounced minima reflect the phase-separated shell of patchy cylinders
(see Fig.~\ref{fig2}(b)) while the homogeneous PE core cylinder does 
not lead to a minimum in this scattering vector regime. 

The dashed lines in Fig.~\ref{fig4}(a) show the calculated results for noninteracting 
patchy cylinders using the same model parameters as in Fig.~\ref{fig3}(b). Some 
features of the measured scattering intensities such as the disappearance of the 
minimum upon increasing the scattering length density of the solvent are captured by 
the theoretical approach. However, the calculated results for noninteracting 
patchy cylinders and the experimental data deviate due to interactions between the micelles. 
The concentration 10 g/L is ten times higher than the one used for the TEM micrographs 
shown in Fig.~\ref{fig1}. This rather high polymer concentration was necessary in order 
to obtain a good signal-to-noise ratio. We emphasize that neither modeling the cores of the 
elongated micelles as one-dimensionally connected objects such as worm-like or 
helical-like chains nor taking into account the semicrystallinity of PE \cite{webe:07} 
leads to a peak of the scattering intensities at \mbox{$q \approx 0.1$ nm$^{-1}$} in the 
cases of dominating contributions of PE to the scattering intensities (upper three 
data sets in Fig.~\ref{fig4}(a)). In order to justify the argumentation based on our 
additional calculations of various intramolecular correlation functions, we have 
performed a scattering experiment for an even higher micelle concentration 
40 g/L in fully deuterated THF (solid squares in Fig.~\ref{fig4}(b)). Indeed the 
observed peak of the scattering intensity is more pronounced as compared to the one 
for 10 g/L (open circles in Fig.~\ref{fig4}(b)) due to 
increased intermolecular correlations. This interpretation is consistent with the observation 
that the 40 g/L sample was rather viscous whereas the 10 g/L sample exhibited fluid-like 
properties. Moreover, a third sample containing 50 g/L triblock terpolymer micelles 
formed a free-standing gel in a simple test tube inversion experiment due to enhanced 
intermolecular correlations.

In order to understand the liquid structure in more detail, the particle-averaged 
total correlation function $h_{OP}({\bf q})$ defined in eqn (\ref{eq5}) has been 
calculated using the polymer reference interaction site integral equation theory 
(see Ref.~\cite{harn:08} and references therein). In contrast to the successful 
application of this theoretical approach to various polymer nanoparticles, 
\cite{henz:08,roch:11,henz:11,webe:07,boli:09} it was not possible to achieve such good 
agreement between the integral equation theory for the rigid patchy cylinders 
shown in Fig.~\ref{fig2}(b) and the experimental data sets across the entire 
$q$-range and for all scattering length contrasts given by the symbols in 
Fig.~\ref{fig4}(a). The differences between the measured and calculated results 
may be due to a number of possible factors with the most critical being the 
model assumption that the PS and PMMA polymer chains form a rigid biphasic shell. 
Due to molecular flexibility, the contribution to the scattering features from 
intermolecular PS and PMMA correlations is less pronounced than the corresponding 
one of a rigid biphasic shell.

Having observed that the integral equation theory for the initial rigid patchy 
cylinders did not lead to a good description of all available scattering data, 
only contributions to the particle-averaged total correlation function arising 
from patchy cylinders with an effective radius \mbox{$R_{eff}=11.5$ nm} 
were considered. The calculated scattering intensities from this model (solid 
lines in Fig.~\ref{fig4}(a)) are comparable to the experimental data. Notably, 
little overall contribution to the scattering features was observed from the total 
correlation function in the case of the lowest scattering length density of 
the solvent (lowest data set in Fig.~\ref{fig4}(a)), due to the presence of 
both positive and negative contributions of the compartments of the triblock 
terpolymer micelles. The peak at \mbox{$q \approx 0.1 $ nm$^{-1}$} observed 
for the three upper data sets in Fig.~\ref{fig4}(a) can be interpreted as a 
sign of intermediate range order with a characteristic real space distance 
of \mbox{$2\pi/q \approx 63$ nm}. Future theoretical work may focus on a detailed 
understanding of local order in fluids consisting of patchy cylinders.

Finally, we note that the experimentally determined scattering intensities shown 
in Fig.~\ref{fig4}(a) can be split consistently into three parts according to 
(see Ref.~\cite{rose:09} and references therein)
\begin{eqnarray}  \label{eq9}
I_{coh}(q)&=&\left({\tilde b}-b_{solvent}\right)^2 I_S(q)+
2\left({\tilde b}-b_{solvent}\right) I_{SI}(q)\nonumber
\\&+&I_I(q)\,,
\end{eqnarray}
where  ${\tilde b} \approx 3.2 \times 10^{10}$ cm$^{-2}$ is the average scattering 
length density of the triblock terpolymer micelles. The first term $I_S(q)$ is the 
normalized scattering intensity of chemically homogeneous micelles, while $I_I(q)$ is 
related to the scattering length inhomogeneity of the triblock terpolymer micelles. The 
cross term $I_{SI}(q)$ between the former contributions can take negative values. In 
general all three terms can be extracted from experimental data if scattering intensities 
have been measured at least at three different scattering length densities of the solvent. 
For the triblock terpolymer micelles under consideration the scattering length density 
independent term $I_I(q)$ is very similar to the middle data set in Fig.~\ref{fig4}(a). 
Moreover, $I_S(q)$ can be considered as the normalized scattering intensity measured at 
infinite contrast, where the last two terms in eqn (\ref{eq9}) can be neglected. The 
functional shape of $I_S(q)$ is similar to the upper data set in Fig.~\ref{fig4}(a) 
for $q \lesssim 0.1$ nm$^{-1}$.

\section{Conclusion}

SANS data have been collected for 
polystyrene-$block$-polyethylene-$block$-poly(methyl methacrylate) triblock terpolymer 
micelles in organic solvents. The structure of these micelles dissolved in protonated 
tetrahydrofuran has been elucidated by comparing the experimentally determined 
scattering intensity to the calculated one for a patchy cylinder (see Fig.~\ref{fig3}(b)). 
Moreover, the theoretical analysis revealed that SANS allows one to distinguish 
patchy cylinders from Janus cylinders (see Fig.~\ref{fig3}(a)). The combined experimental 
and theoretical study shows the presence of alternating polystyrene and 
poly(methyl methacrylate) regions in the shell of the patchy, cylindrical triblock 
terpolymer micelles (see Fig.~\ref{fig2}(b)).

It has not been possible to use the polymer reference integral equation theory for rigid 
patchy cylinders to interrogate the scattering data of the triblock terpolymer micelles 
at rather high concentration in various mixtures of protonated and deuterated tetrahydrofuran. 
The principle reason for this may be that the amorphous polystyrene and poly(methyl methacrylate) 
chains forming the shell of the core-shell micelles require chain flexibility to be taken 
into account. However, consistent information was obtained by separating the data analysis 
into two portions, analyzing the scattering data considering noninteracting patchy 
cylinders (see dashed lines in Fig.~\ref{fig4}(a)) and taking into account intermolecular 
correlations from patchy cylinders with an effective shell radius (see solid lines in 
Fig.~\ref{fig4}(a)). Upon further increasing the micelle concentration the SANS data show, 
categorically, the presence of pronounced intermolecular correlations (see Fig.~\ref{fig4}(b)).

\section*{Acknowledgement}

This work was 
supported by the German Science Foundation (Collaborative Research Center 840, project A2). 
We thank the J\"ulich Center of Neutron Scattering (JCNS, Germany) for financial travel support 
and for providing beamtime at the KWS 1 instrument at FRM II in Munich, Germany.


\begin{thebibliography}{99}

\bibitem{foe97} S. F\"orster,
{\it Ber. Bunsenges. Phys. Chem.}, 1997, {\bf 101}, 1671.

\bibitem{bat99} F. S. Bates and G. H. Frederickson,
{\it Physics Today}, 1999, {\bf 52}, 32.

\bibitem{ham01} I. W. Hamley, S.-M. Mai, A. J. Ryan, J.~P.~A. Fairclough and C. Booth,
{\it Phys. Chem. Chem. Phys.}, 2001, {\bf 3}, 2972.

\bibitem{hel11} T. Hellweg, in {\it Self Organized Nanostructures of Amphiphilic Block Copolymers II}, 
ed. O. Borisov and A. H. E. M\"uller,
Springer, Heidelberg, 2012, pg. 1.

\bibitem{ruz05} A.-V. Ruzette and L. Leibler,
{\it Nature Mater.}, 2005, {\bf 4}, 19.

\bibitem{rod05} J. Rodriguez-Hernandez, F. Checot, Y. Gnanou and S. Lecommandoux,
{\it Prog. Polym. Sci.}, 2005, {\bf 30}, 691.

\bibitem{hab09} N. Haberkorn, M. C. Lechmann, B. H. Son, K. Char, J. S. Gutmann and P. Theato,
{\it Macromol. Rapid Comm.}, 2009, {\bf 30}, 1146.

\bibitem{mot10} M. Motornov, Y. Roiter, I. Tokarev and S. Minko,
{\it Prog. Polym. Sci.}, 2010, {\bf 35}, 174.

\bibitem{kim10a} J. K. Kim, S. Y. Yang, Y. Lee and Y. Kim,
{\it Prog. Polym. Sci.}, 2010, {\bf 35}, 1325.

\bibitem{kim10} H.-C. Kim, S.-M. Park and W. D. Hinsberg,
{\it Chem. Rev.}, 2010, {\bf 110}, 146.

\bibitem{had05} N. Hadjichristidis, H. Iatrou, M. Pitsikalis, S. Pispas and A. Avgeropoulos,
{\it Progr. Polym. Sci.}, 2005, {\bf 30}, 725.

\bibitem{dup10} J. Dupont and G. Liu,
{\it Soft Matter}, 2010, {\bf 6}, 3654.

\bibitem{du11} J. Z. Du and R. K. O'Reilly,
{\it Chem. Soc. Rev.}, 2011, {\bf 40}, 2402.

\bibitem{zha12} K. Zhang, M. Jiang, D. Chen, {\it Prog. Polym. Sci.}, 2012, {\bf 37}, 445.

\bibitem{mou12} A. O. Moughton, M. A. Hillmeyer, T. P. Lodge, {\it Macromolecules}, 2012, {\bf 45}, 2.

\bibitem{wal08a} A. Walther and A. H. E. M\"uller,
{\it Soft Matter}, 2008, {\bf 4}, 663.

\bibitem{wal08} A. Walther, K. Matussek and A. H. E. M\"uller,
{\it ACS Nano}, 2008, {\bf 2}, 1167.

\bibitem{voe:09} I. K. Voets, R. Fokkink, T. Hellweg, S. M. King, P. de Waard, A. de Keizer, M. A. Cohen Stuart, {\it 
Soft Matter}, 2009, {\bf 5}, 999.

\bibitem{jia10} S. Jiang, Q. Chen, M. Tripathy, E. Luijten, K. S. Schweizer and S. Granick,
{\it Adv. Mater.}, 2010, {\bf 22}, 1060.

\bibitem{kre11} I. Kretzschmar and J. H. Song,
{\it Curr. Opinion Coll. Interf. Sci.}, 2011, {\bf 16}, 84.

\bibitem{erh01} R. Erhardt, A. B\"oker, H. Zettl, H. Kaya, W. Pyckhout-Hintzen, G. Krausch,
V. Abetz and A. H. E. M\"uller,
{\it Macromolecules}, 2001, {\bf 34}, 1069.

\bibitem{glo07} S. C. Glotzer and M. J. Solomon,
{\it Nature Mater.}, 2007, {\bf 6}, 557.

\bibitem{paw10} A. B. Pawar and I. Kretzschmar,
{\it Macromol. Rapid Comm.}, 2010, {\bf 31}, 150.

\bibitem{liu03} Y. Liu, V. Abetz and A. H. E. M\"uller,
{\it Macromolecules}, 2003, {\bf 36}, 7894.

\bibitem{ruh11} T. M. Ruhland, A. H. Gr\"oschel, A. Wather and A. H. E. M\"uller,
{\it Langmuir}, 2011, {\bf 11}, 9807

\bibitem{the10} P. E. Theodorakis, W.  Paul and K. Binder,
{\it Macromolecules}, 2010, {\bf 43}, 5137.

\bibitem{eru11} I. Erukhimovich, P. E. Theodorakis, W.  Paul and K. Binder,
{\it J. Chem. Phys.}, 2011, {\bf 134}, 054906.

\bibitem{the11} P. E. Theodorakis, H.-S. Hsu, W.  Paul and K. Binder,
{\it J. Chem. Phys.}, 2011, {\bf 135}, 164903.

\bibitem{laz09} M. Lazzari and A. Lopez-Quintela,
{\it Macromol. Rapid Comm.}, 2009, {\bf 21}, 1785.

\bibitem{dup09} J. Dupont, G. J. Liu, K. Niihara, R. Kimoto and H. Jinnai,
{\it Angew. Chem. Int. Ed.}, 2009, {\bf 48}, 6144.

\bibitem{wan07} X. Wang, G. Guerin, H. Wang, Y. Wang, I. Manners and M. A. Winnik,
{\it Science}, 2007, {\bf 317}, 644.

\bibitem{gae09} T. Gaedt, N. S. Ieong, G. Cambridge, M. A. Winnik and I. Manners,
{\it Nature Mater.}, 2009, {\bf 8}, 144.

\bibitem{he11} F. He, T. Gaedt, I. Manners and M. A.  Winnik,
{\it J. Am. Chem. Soc.}, 2011, {\bf 133}, 23.

\bibitem{qia10} J. Qian, M. Zhang, I. Manners and M. A. Winnik,
{\it Trends Biotechnol.}, 2010, {\bf 28}, 84.

\bibitem{schm:08c} H. Schmalz, J. Schmelz, M. Drechsler, J. Yuan, A. Walther, K. Schweimer
 and A. M. Mihut,
{\it Macromolecules}, 2008, {\bf 41}, 3235.

\bibitem{schm:11} J. Schmelz, M. Karg, T. Hellweg and H. Schmalz,
{\it ACS Nano}, 2011, {\bf 5}, 9523.

\bibitem{fue04} T. Fuetterer, A. Nordskog, T. Hellweg, G. H. Findenegg , S. Foerster and C. D. Dewhurst,
{\it Phys. Rev. E}, 2004, {\bf 70}, 041408.

\bibitem{walt:09} A. Walther, M. Drechsler, S. Rosenfeldt, L. Harnau, M. Ballauff, 
V. Abetz and A. H. E. M\"uller, 
{\it J. Am. Chem. Soc.}, 2009, {\bf 13}, 4720.

\bibitem{will:91} C. E. Williams in {\it Neutron, X-Ray and Light Scattering}, 
ed. P. Lindner and T. Zemb,
Elsevier Science Publisher, Oxford, 1991, pg. 101.

\bibitem{bals:94} N. P. Balsara, D. J. Lohse, W. W. Graessley and R. Krishnamoorti,
{\it J. Chem. Phys.}, 1994, {\bf 100}, 3905.

\bibitem{higg:96} {\it Polymers and Neutron Scattering}, J. S. Higgins and H. C. Benoit,
Clarendon Press Ithaca, Oxford, 1996.

\bibitem{rose:09} S. Rosenfeldt, M. Ballauff, P. Lindner and L. Harnau,
{\it J. Chem. Phys.}, 2009, {\bf 130}, 244901.

\bibitem{yan:08} Y. Yan, L. Harnau, N. A. M. Besseling, A. de Keizer, M. Ballauff, S. Rosenfeldt 
and M. A. Cohen-Stuart, 
{\it Soft Matter}, 2008, {\bf 4}, 2207.

\bibitem{henz:08} K. Henzler, S. Rosenfeldt, A. Wittemann, L. Harnau, S. Finet, T. Narayanan 
and M. Ballauff, 
{\it Phys. Rev. Lett.}, 2008, {\bf 100}, 158301.

\bibitem{roch:11} C. N. Rochette, S. Rosenfeldt, K. Henzler, F. Polzer, M. Ballauff, Q. Tong, 
S. Mecking, M. Drechsler, T. Narayanan and L. Harnau, 
{\it Macromolecules}, 2011, {\bf 44}, 4845.

\bibitem{henz:11} K. Henzler, B. Haupt, S. Rosenfeldt, L. Harnau, T. Narayanan and M. Ballauff, 
{\it Phys. Chem. Chem. Phys.}, 2011, {\bf 13}, 17599.

\bibitem{webe:07} C. H. M. Weber, A. Chiche, G. Krausch, S. Rosenfeldt, M. Ballauff, L. Harnau, 
I. G\"ottker-Schnetmann, Q. Tong and S. Mecking, 
{\it Nano Lett.} 2007, {\bf 13}, 2024.

\bibitem{harn:08} L. Harnau, 
{\it Mol. Phys.}, 2008, {\bf 106}, 1975.

\bibitem{boli:09} S. Bolisetty, S. Rosenfeldt, C. Rochette, L. Harnau, P. Lindner, Y. Xu, 
A. H. E. M\"uller and M. Ballauff, 
{\it Colloid Polym. Sci.} 2009 {\bf 287}, 129.

\end{thebibliography}
\end{document}